\documentclass[pre,amsmath,aps,superscriptaddress,twocolumn]{revtex4}
\usepackage{graphicx,xspace}
\usepackage{float}
\usepackage{amsmath}
\usepackage{amssymb}
\usepackage[normalem]{ulem}
\usepackage{epsfig}
\usepackage{graphicx,psfrag,xspace}

\begin{document}

\title{Mass transport subject to time-dependent flow with non-uniform sorption in porous media}
\author{Marie-Christine N\'eel}
\email{mcneel@avignon.inra.fr}
\affiliation{Universit\'e d'Avignon et des Pays de Vaucluse, UMR 1114 EMMAH, 84018 Avignon Cedex, France}
\author{Andrea Zoia}
\email{andrea.zoia@cea.fr}
\affiliation{CEA/Saclay, DEN/DM2S/SFME/LSET, B\^at.~454, 91191 Gif-sur-Yvette Cedex, France}
\author{Maminirina Joelson}
\email{maminirina.joelson@univ-avignon.fr}
\affiliation{Universit\'e d'Avignon et des Pays de Vaucluse, UMR 1114 EMMAH, 84018 Avignon Cedex, France}

\begin{abstract}
We address the description of solutes flow with trapping processes in porous media. Starting from a small-scale model for tracer particles trajectories, we derive the corresponding governing equations for the concentration of the mobile and immobile phases. We show that this formulation is fairly general and can easily take into account non-constant coefficients and in particular space-dependent sorption rates. The transport equations are solved numerically and a comparison with Monte Carlo particle-tracking simulations of spatial contaminant profiles and breakthrough curves is proposed, so to illustrate the obtained results.
\end{abstract}

\maketitle

\section{Introduction}

Contaminant migration in porous media is often characterized by non-Fickian (anomalous) transport: following injection, the spread of the pollutants plume might grow nonlinearly in time, $\langle x^2(t) - \langle x(t) \rangle^2  \rangle \sim t^\gamma$, $\gamma \ne 1$, and the resulting concentration profiles display a non-Gaussian behavior~\cite{sahimi, scher_framework, rev_geo}. In contrast, particles flow in perfectly homogeneous media (where the Fickian advection-dispersion mechanisms apply) gives rise to linear spread and Gaussian shapes: see, e.g.,~\cite{cortis_homog} and references therein. Many concurrent processes may explain the observed deviations from Gaussianity. For instance, the presence of irregularities at multiple space scales~\cite{fractures, levy}, the complex structures of flow streams~\cite{kirchner, zoia} and saturation/stagnation distribution within the medium~\cite{bromly}, and the physical-chemical or bio-physical exchanges of the pollutant particles with the surrounding material~\cite{berkowitz_sorp, Tufen} make the homogeneity hypothesis questionable.

Anomalous transport often displays non-universal features: different physical conditions lead to concentration profiles that, while sharing some properties (such as the scaling law for the spread, for instance), can not be interpreted within a single coherent framework. This is especially apparent in presence of boundaries~\cite{zoia_prl}, and explains the coexistence of several models aimed at understanding and predicting solutes dynamics in complex materials. Among them, some of the most widely adopted formulations are the Continuous Time Random Walk (CTRW)~\cite{rev_geo, scher_framework} and the fractal Mobile-Immobile Model (f-MIM)~\cite{Haggerty, Schumer}: both have indeed been applied with success to the analysis of experimental data ranging from laboratory to field scale~\cite{rev_geo, Boano, benson_m3, levy}. In particular, these approaches are well suited to shed light on the `heavy tailed' (power-law decaying) breakthrough curves (BTCs) that are frequently measured at the outlet of experimental setups, and the non-Gaussian shapes of spatial contaminant profiles. For a detailed discussion on the distinct features, advantages and limitations of these models see, e.g.,~\cite{rev_geo, Schumer, benson_m3, Zhang2, berkowitz_sorp}.

The long-time asymptotic behavior of these formulations may look very similar (up to an appropriate renaming of the coefficients)~\cite{rev_geo} when the analysis is limited to given physical quantities (such as BTCs), whereas relevant discrepancies may appear at a closer inspection (by examining, e.g., spatial profiles as we will see in the following). Moreover, the governing equations for anomalous transport sometimes appear in the literature under different forms, equivalent when all parameters, e.g., the dispersion coefficient or the velocity, are constant and uniform. This equivalence may break down when the parameters depend on position and/or time (see, e.g.,~\cite{Heinsa}). In all such cases, one must adopt more precise hypotheses on the microscopic solute dynamics in the traversed medium, so to single out an appropriate model for the experimental data under consideration. In general, however, solutes trajectories are hardly accessible by experiments (at least in the context of contaminant transport in porous media), so that one must resort to ensemble-averaged macroscopic measurable quantities in order to discriminate between hypotheses. For instance, the comparison between model-predicted and experimentally measured BTCs and spatial profiles may allow choosing the most appropriate conceptual framework.

According to the f-MIM model, in the hydrodynamic limit the evolution of the contaminant plume density is ruled by a transport equation involving integral-differential operators of non-integer (fractional) order in time, as shown in~\cite{Zhang2, Mary} on the basis of numerical and theoretical arguments. Here the hydrodynamic limit refers to the fact that we are observing the plume behavior at space and time scales much longer than those characterizing the typical particles displacements. The prototype equation for the evolution of so-called fractional dynamics is the Fractional Fokker-Planck Equation (FFPE)~\cite{Zaslavsky, Barkai}: indeed, f-MIM and FFPE share many features, and show a similar asymptotic behavior. However, these two equations are not equivalent, and represent hydrodynamic limits of distinct small-scale models for particles trajectories.

In particular, the FFPE corresponds to random walks performing Gaussian jumps (in potential fields) that take random durations to be completed~\cite{Barkai, Madziarz, MetK2}, whereas f-MIM describes random walks involving immobile periods (of random duration) and Gaussian displacements at each time step~\cite{VieVG, Schumer, Zhang2, benson_m3}. Conceptually, the former is an expedient means of describing a broad spectrum of velocities, such those characterizing flows in heterogeneous and/or nonsaturated media, whereas the latter allows distinguishing between a solid matrix (where particles are stuck, such as in a low permeability region: the so-called immobile phase) and a bulk flow (where particles undergo advection and dispersion processes: the so-called mobile phase).

The aim of our work is to provide a generalization of the f-MIM approach to the case of non-constant flows, and space-dependent sorption rates, which can commonly arise in transport experiments. Non-uniformity can occur in both field and laboratory scale measures, and may involve also sharp discontinuities in the physical properties of the traversed media. Based on a small-scale description of particles trajectories, we will derive the corresponding governing equations for the macroscopic quantities, namely the mobile and immobile densities, in such a way that non-constant coefficients are easily taken into account. In the case of constant velocity field and uniform sorption, similar results were obtained via subordination theory~\cite{Mary}, along the strategy proposed in~\cite{Barkai2, Madziarz, Pirya,GorMaiVi}. In the case of non-constant coefficients, stronger arguments have to be used. Finally, in order to corroborate the proposed results, we will compare Monte Carlo particle-tracking simulations of solutes spatial profiles and BTCs with the numerical solutions of the governing equations.

This paper is organized as follows: in Sec.~\ref{mim_model} we recall the small-scale model for flow with trapping processes in porous media, on the basis of the f-MIM formalism, and provide an extension to non-constant flows. Then, in Sec.~\ref{densities}, we derive the densities of the mobile and immobile contaminant phases at small scale, and in Sec.~\ref{equations} the corresponding governing equations in the hydrodynamic limit. These equations are discretized and solved numerically, and the obtained solutions are compared with Monte Carlo simulations results in Sec.~\ref{montecarlo}. Conclusions are finally drawn in Sec.~\ref{conclusions}.

\section{A small-scale model for flow with trapping processes}
\label{mim_model}

As customary, we begin by representing the stochastic trajectory of a contaminant particle in a porous medium as a random walk $x_t^{\ell,\tau}$ undergoing advection and dispersion. Superscripts $\ell,\tau$ denote the characteristic length and time scales, respectively, of the process. First, we briefly recall the essential features of the standard Gaussian models that usually describe small-scale displacements of a contaminant plume in homogeneous saturated materials. Then, we focus on heterogeneous and/or unsaturated media, where the broad distribution of permeabilities and different flow regions experienced by the tracers is mirrored in the possibility of trapping events at each visited spatial site, the walkers dynamics being otherwise similar to that observed in homogeneous materials. These trapping events affect the sojourn times and thus alter the typical scales of average displacement (i.e., velocity) and spread (i.e., dispersion) of the contaminant plume.

\subsection{A Gaussian model for homogeneous materials}

For homogeneous saturated materials, it is usually possible to identify an average flow field $v(t)$, so that at each time step $[t-\tau,t]$ particles are advected over a distance $ \mu(t)=\int_{t- \tau}^t v(t')dt'$. Dispersion is usually taken into account by adding random (symmetrical) jumps of characteristic length scale $\ell$ to the advective contribution $\mu$. The total displacement during $\tau$ can be therefore written as $\Delta x=  \mu + \ell \xi$, where $\xi$ is a random number drawn from a probability density function (pdf) $\varphi_1(\xi)$ with zero mean and unit variance. It follows that $\varphi_\ell(\cdot)= \ell^{-1} \varphi_1(\cdot/ \ell)$ is the density of $\ell \xi$. Usually, one assumes that $\varphi_\ell(\xi)$ is a normal pdf with zero mean and standard deviation equal to $\ell$, which means that the typical scale of fluctuations around the average particle displacement is $\ell$. It is well known that the scaling (hydrodynamic) limit of such (independent) random walks is attained when $\tau \to 0$ in such a way that the ratio $\ell^2/2\tau$ converges to some limit $D$. The parameter $D$ defines the dispersion coefficient. Under this scaling limit and when $v$ is constant, the stochastic process $x_t^{\ell,\tau}$ asymptotically approaches the Brownian Motion (BM) $x_t$ with drift, whose concentration $P(x,t)$ (i.e., the probability density of finding a walker at position $x$ at a given time $t$) is shown to satisfy the following Fokker-Planck Equation (FPE)~\cite{Risken}
\begin{equation}
\partial_t P(x,t)=\partial^2_x D P(x,t) - \partial_x v P(x,t).
\label{ade}
\end{equation}
Note that the FPE~\eqref{ade} is equivalent to the Advection-Dispersion Equation (ADE) when $D$ is uniform~\cite{delay}. When $D$ depends on $x$, Eq.~\eqref{ade} still represents the hydrodynamic limit of random walks as above, satisfying $\ell^2/2\tau=D(x)$ ~\cite{delay}. For sake of simplicity, in the following we will refer to the case where $D$ is uniform, i.e., $D(x)=D$. By virtue of the Central Limit Theorem, the results recalled above actually apply more generally to a broad class of random walks where $\varphi_\ell(\xi)$ is a generic symmetric jump length pdf with finite second moment. After a sufficient number of displacements, these processes all converge to BM in the hydrodynamic limit (provided that the variance of the process is equal to $\ell^2$, and with constant $v$). The fact that BM is the basin of attraction of a large spectrum of random walks, independently of the specific choice of the jump length pdf, can explain the success of Eq.~\eqref{ade} in interpreting experimental contaminant transport data, at least limited to homogeneous saturated materials~\cite{cortis_homog}. Remark also that Eq.~\eqref{ade} can be derived by building upon a constitutive relationship for the particles flux (probability current) ${\mathcal F}^{\ell,\tau}(x,t)$ of the process $x_t^{\ell,\tau}$, as recalled in Appendix~\ref{D}. In the hydrodynamic limit, the flux converges to
\begin{equation}
{\mathcal F}(x,t)=v P(x,t)-\partial_x D P (x,t),
\label{fick}
\end{equation}
which is the well-known Fick's law for homogeneous $D$. Then, mass conservation principle $\partial_t P(x,t) = -\partial_x {\mathcal F}(x,t)$ yields Eq.~\eqref{ade}. This holds for infinite domains, or for domains limited by absorbing boundary conditions (walkers are removed upon touching the barriers). For detailed accounts concerning the derivation of Eq.~\eqref{ade}, see, e.g.,~\cite{MetK2,delay, Hugues}.

\subsection{Dispersion with trapping events}

Suppose now that the traversed material is affected by small-scale heterogeneities. For instance, we might consider unsaturated porous media with variable saturation, where particles may be retained by stagnation regions~\cite{bromly}. We might also think of biophysical effects in porous materials~\cite{Tufen}, or chemical reactions taking place on the surface of a duct traversed by fluid flow. The homogeneous random walk proposed above is evidently inadequate to address such situations. From the point of view of microscopic trajectories, a natural means of accounting for sorption is that walkers are given the possibility of being trapped at the end of each displacement, i.e., upon reaching a new spatial site. The trapping probability $h(x)$ is for sake of generality space-dependent, since traps may be non-uniformly distributed. Previously,~\cite{Madziarz, Mary, Zhang2} considered constant trapping probabilities. We further assume that the sojourn time at the traps is itself a random variable $t_w$. Experimental evidences suggest that these retention times lack a characteristic scale (i.e., their average is not defined), so that it is commonly assumed that $t_w$ obeys a power-law decaying pdf~\cite{Schumer, bromly}.

We introduce the scaled variable $t_w = \tau^{1/\gamma} W$, where $W$ is a random number obeying a pdf $\psi$ and $\gamma$ a scaling exponent, and denote by $\Psi$ the associated survival probability, $\Psi(t)=\int_t^{+\infty}\psi(t')dt'$. The quantity $\Psi(t)$ expresses the probability that the trapping time is longer than $t$. It follows that the rescaled pdf of $t_w$ is $\psi_\gamma=\tau^{-1/\gamma}\psi(t/\tau^{1/\gamma})$, and the rescaled survival probability is $\Psi_\gamma=\Psi(t/\tau^{1/\gamma})$. We make now the following hypothesis for the pdf of the retention times:\\

$H_1$) the pdf $\psi$ is concentrated on $R^+$, with survival probability of the kind $\Psi(t)=\lambda t^{-\gamma}/\Gamma(1-\gamma)+ K(t)$, $K$ being a function integrable over $R^+$, with $0<\gamma<1$.\\

Assumption $H_1$ is satisfied by pdfs whose asymptotic behavior is a power-law. For instance, we might consider Pareto laws~\cite{zaliapin, zoia_pareto}, or maximally skewed L\'evy laws with exponent $\gamma$, which are concentrated on $R^+$ precisely for $0<\gamma<1$~\cite{Fel, GneK, Lev, MeerSchef}. Intuitively, the exponent $\gamma$ quantifies the degree of heterogeneity of the porous media: small values of $\gamma$ denote strong deviations from the usual Gaussian transport model, i.e., anomalously long retention times.

While the retention times correspond to the immobilization of the walkers (immobile phase), a more precise description of the time spent during displacements (mobile phase) is needed. Several scenarios can be conceived, depending on the time of occurrence of the dispersive jump within a mobile period $[t-\tau, t]$. For instance, the jump could take place instantaneously at the beginning of the period, or at the end; or it could occur at a random time, uniformly distributed in $[t-\tau,t]$. Also, we could imagine that the jump is not instantaneous, and takes the whole time span $[t-\tau,t]$ to be completed. On the other hand, the endpoints of successive displacements do not depend on the considered scenario, nor are trapping events affected. As shown in Appendix~\ref{A}, all these possible small-scale random walks converge to the same scaling limit when $\tau, \ell \to 0$. Then, for convenience we will focus on the simplest case: we assume that walkers perform a single instantaneous dispersive jump during the time interval $\tau$, taking place at the end of each mobile period.

Finally, for sake of generality, we also introduce a (possibly time- and space-dependent) source term $r(x,t)$, representing tracer injection.

In the following, we will show that in the hydrodynamic limit the walkers density $P(x,t)$ for the process described above satisfies
\begin{equation}
\partial_t P=\partial^2_x D{\mathcal H}_{\lambda,\gamma,h} P- \partial_x v {\mathcal H}_{\lambda,\gamma,h} P+r.
\label{eqmim}
\end{equation}
In Eq.~\eqref{eqmim}, the non-local in time operator ${\mathcal H}_{\lambda,\gamma,h}$ is the inverse of the (also non-local in time) mapping $\text{Id}+\lambda h(x)I_{0,+}^{1-\gamma}$, which entails the fractional integral of order $1-\gamma$, namely $I_{0,+}^{1-\gamma}$, whose definition is recalled in Appendix~\ref{B}. Here $\text{Id}$ denotes the identity operator and $\lambda \ge0 $ is a constant parameter. In fact, as shown in~\cite{Mary}, ${\mathcal H}_{\lambda,\gamma,h}$ is the time convolution of the kernel $\frac{d}{dt}E_{1-\gamma}[-\lambda h(x) t^{1-\gamma}]$, where $E_\alpha$ is the Mittag-Leffler function described in~\cite{MaiUd, MaiGorAna, Kilb}. In~\cite{Mary} it was shown that Eq.~\eqref{eqmim} governs the evolution of the particles concentration for constant $v$ and uniform $D$, with $h(x)\equiv 1$, building upon the results of~\cite{Zhang2}; for this case, and assuming $r(x,t)=\delta(t)\delta(x)$, Eq.~\eqref{eqmim} is equivalent to the fractal MIM model
\begin{equation}
\left(\partial_t+\lambda \partial_t^\gamma \right) P(x,t)=\partial_x (D\partial_x-v)P(x,t)
\label{derqmim}
\end{equation}
introduced in~\cite{Schumer}, where $\partial_t^\gamma$ is the Caputo derivative of order $\gamma$ (see Appendix~\ref{B}). The exponent $\gamma<1$ characterizes the asymptotic behavior of the trapping times pdf (in $H_1$), and also the scaling of the plume spread; in this sense, $\gamma$ is the signature of the anomalous transport process. Indeed, the solutions of Eq.~\eqref{derqmim} have been shown to decrease at large times as  $t^{-\gamma}$~\cite{Schumer}, which could possibly explain the long tails experimentally observed, e.g., by~\cite{bromly} and~\cite{kirchner}. In unbounded domains, and with constant and uniform coefficients, the spatial moments of the solute concentration in Eq.~\eqref{derqmim} were shown to decrease as powers of time related to the exponent $\gamma$~\cite{Zhang2}.

As a special limit, setting $\gamma=1$ in Eq.~\eqref{derqmim} yields the well-known MIM model with retardation factor $1+\lambda$~\cite{VieVG}, which corresponds to tracers experiencing random retention periods with finite characteristic (mean) duration, comparable to the time spent in the average flow field. Nevertheless, the solutions of the MIM model or the ADE~\eqref{ade} fall off much more rapidly than any power of $t$ and are thus inadequate to interpret experimental data showing heavy tails such as those of~\cite{bromly, kirchner}. We will see further below that, rather than  representing the hydrodynamic limit of random walks satisfying $H_1$ with $\gamma=1$, the MIM model corresponds to a pdf $\psi$ with a finite average.

The parameter $\lambda$ determines the relevance of the retention mechanism with respect to the Fickian transport: when $\lambda=0$, all equations above collapse to the ADE. Moreover, it provides the scaling parameter for $\psi$, and carries dimensions of a power $\gamma$ of time. This is easily seen in the Laplace space, where $\psi(s) \simeq 1-\lambda s^\gamma$ in the limit $s \to 0$, according to $H_1$.

\section{Probability densities for the mobile and immobile phases at small scale}
\label{densities}

Particles performing such random walks with sorption can conceptually be separated into two distinct `phases': at each time step, walkers that are trapped are said to be in the immobile phase, whereas walkers that are not are said to be in the mobile phase. In the following, we proceed to derive an explicit relation that links the particles densities in the two phases, for definite values of length- and time-scales $\ell$ and $\tau$. The hydrodynamic limit will be addressed in next Section.

Let $P_i^{\ell,\tau}(x,t)$ be the density of trapped particles, at location $x$ at time $t$, and $P_m^{\ell,\tau}(x,t)$ the density of mobile walkers. In order to establish the desired relation between the two spatial densities $P_i^{\ell,\tau}$ and $P_m^{\ell,\tau}$, we make use of the ancillary pdfs $p_j^{\ell,\tau}$ of just arriving at point $x$ at time $t$, and $p_m^{\ell,\tau}$ of just being released by a sorbing site at time $t$.

Except just after having been injected into the system, mobile particles at position $x$ at time $t$ have two alternatives. Either they may have completed a mobile period at time $t-t'$ ($0<t'<\tau$), without being trapped; or, they may have been trapped and then released, at a distance $\int_{t-t'}^tv(\theta)d\theta$ from $x$. Both possible events are followed by a convective displacement that may not be completed at time $t$. The displacement completed at time $t$ has amplitude $L(t,t')=\int_{t-t'}^tv(\theta )d\theta$. Remark that $L(t,t')=vt'$ if $v$ is constant. Hence, we have the following probability balance
\begin{equation}
P_m^{\ell,\tau}(x,t)=\int_0^\tau {\mathcal T}_{t'}{\mathcal Y}_{L(t,t')}[ f^{\ell,\tau}+r](x,t)dt',
\label{pmlt}
\end{equation}
where the quantity
\begin{equation}
f^{\ell,\tau}(x,t)=p_m^{\ell,\tau}(x,t)+\left[ 1-h(x) \right] p_j^{\ell,\tau}(x,t)
\end{equation}
is the pdf of just beginning a mobile period at time $t$ and position $x$, after a previous mobile period (i.e., particles just injected by the source are excluded). Convective displacements are represented by means of the operators ${\mathcal T}_{u}$ and ${\mathcal Y}_{w}$, which denote translation in time and space, respectively; i.e., ${\mathcal T}_{u}G(t)=\left[HG\right](t-u)$, $H$ being the Heaviside step function, and ${\mathcal Y}_{w}g(x)=g(x-w)$. Eq.~\eqref{pmlt} corresponds to scenario (S1) of Appendix~\ref{A}, the dispersive jumps occurring at the end of each mobile period.

Immobile particles that are in $x$ at time $t$ must have jumped there previously, been trapped and stayed there up to $t$. Hence, denoting time convolutions of functions in $R^+$ by $*$, i.e., $F*G(t)=\int_0^t F(t-t')G(t')dt'$, we have
\begin{equation}
P_i^{\ell,\tau}(x,t)=h(x) \Psi_\gamma *  p_j^{\ell,\tau}.
\label{pilt}
\end{equation}

To complete the mass balance above we need another equation. Particles just arriving at $x$ at time $t>\tau$ may $i)$ have jumped at the previous time step without being trapped, $ii)$ have been trapped and released, or $iii)$ may have been injected into the system by the source, in each case at time $t-\tau$. Hence, for $ p_j^{\ell,\tau}(x,t)$, which appears on the right-hand side of Eq.~\eqref{pilt}, we have
\begin{equation}
p_j^{\ell,\tau}(x,t)={\mathcal T}_\tau {\mathcal Y}_{L(t,\tau)} [f^{\ell,\tau}+r] \star \varphi_\ell
\label{ajlt}
\end{equation}
for $t>\tau$, where $\star$ denotes space convolution, i.e., $f\star g(x)=\int_Rf(x-x')g(x')dx'$. Note that the explicit dependence of the convolution product on the variables $(x,t)$ has been omitted.

According to Eq.~\eqref{pmlt}, $\tau^{-1}P_m^{\ell,\tau}(x,t)$ is the average of ${\mathcal T}_{t'}{\mathcal Y}_{L(t,t')}[ f^{\ell,\tau}+r](x,t)$ over an interval of amplitude $\tau$, hence approximates ${\mathcal T}_{\tau}{\mathcal Y}_{L(t,\tau)}[ f^{\ell,\tau}+r](x,t)$ when $\tau$ becomes small, at least for smooth functions of time. Since convolutions as in Eqs.~\eqref{pilt} and~\eqref{ajlt} have a smoothing effect, this latter assumption is not necessary to ensure that replacing ${\mathcal T}_\tau {\mathcal Y}_{L(t,\tau)} [f^{\ell,\tau}+r]$ by $\tau^{-1}P_m^{\ell,\tau}$ results into a small error for $P_i^{\ell,\tau}$. To check this argument, let us just split ${\mathcal T}_\tau {\mathcal Y}_{L(t,\tau)} [f^{\ell,\tau}+r]$ into $\tau^{-1}P_m^{\ell,\tau}$ and the remainder, and estimate the influence of this latter in $ p_j^{\ell,\tau}(x,t)$ and Eq.~\eqref{pilt}. We obtain
\begin{equation}
P_i^{\ell,\tau}(x,t)=h(x) \left[ R^{\ell,\tau}P_m^{\ell,\tau}(x,t)+E^{\ell,\tau}(x,t) \right].
\label{imlt}
\end{equation}
Further below, we will address the limit of the operator $R^{\ell,\tau}$, and check that $E^{\ell,\tau}$ tends to zero when $\ell,\tau\to 0$. We have set
\begin{equation}
R^{\ell,\tau}g(x,t)=\tau ^{-1}\left[\Psi_\gamma * g \star \varphi_\ell \right](x,t),
\label{rlt}
\end{equation}
\begin{equation}
E^{\ell,\tau}(x,t)=\Psi_\gamma * \varphi_\ell \star \varepsilon^{\ell,\tau}(x,t),
\label{elt}
\end{equation}
and
\begin{eqnarray}
&\varepsilon^{\ell,\tau}(x,t)={\mathcal T}_{\tau}{\mathcal Y}_{L(t,\tau)} [f^{\ell,\tau}+r](x,t)-\frac{P_m^{\ell,\tau}(x,t)}{\tau}=\nonumber \\
&\int_0^1 ({\mathcal T}_{\tau} {\mathcal Y}_{L(t,\tau)}-{\mathcal T}_{\theta\tau}{\mathcal Y}_{L(t,\theta\tau)}) [f^{\ell,\tau}+r](x,t)d\theta.
\label{epslt}
\end{eqnarray}
The equations above provide the link between the mobile and immobile walkers densities, at small scale.

\section{Governing equations}
\label{equations}

In this Section, we derive the hydrodynamic limit of the small scale processes described above, and illustrate the macroscopic governing equations.
The starting point is the limit of Eq.~\eqref{imlt}.

\begin{figure}[t]
\centerline{\epsfxsize=9.0cm\epsfbox{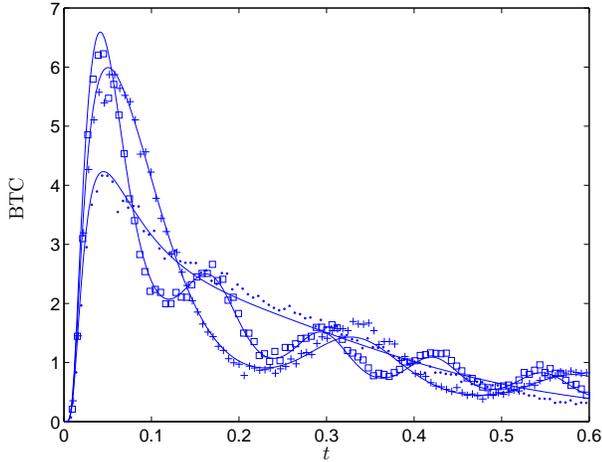}}
\caption{Breakthrough curve ${\mathcal F}(x,t)$ at the right outlet of a domain with length $L=1$. The time-dependent velocity is $v(t)=2 \sin (\nu t)$, with $D=1$, $h\equiv 1$,  $\lambda=1$ and $\gamma=0.5$. A point-source is located in $x_0=L/2$ at time $t=0$. The left outlet has reflective boundary conditions, while the right outlet has absorbing boundary conditions. Symbols represent Monte Carlo simulation (dots $\nu=5$, squares $\nu=25$, and crosses $\nu=50$), solid line numerical integration.}
   \label{fig1}
\end{figure}

\begin{figure}[t]
\centerline{\epsfxsize=9.0cm\epsfbox{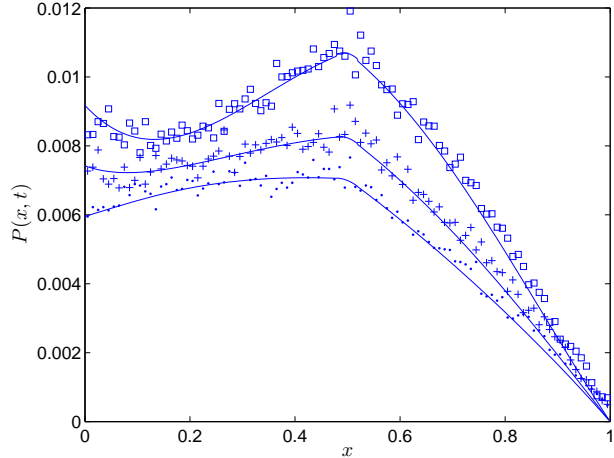}}
\caption{Total concentration profiles $P(x,t)$ in domain with length $L=1$, at fixed times. The time-dependent velocity is $v(t)=2 \sin (\nu t)$, with $D=1$, $h\equiv 1$, $\lambda=1$, $\nu=50$ and $\gamma=0.5$. Initial data and boundary conditions are as in Fig.~\ref{fig1}. Symbols represent Monte Carlo simulation (squares $t=0.1$, crosses $t=0.2$, and dots $t=0.3$), solid lines numerical integration.}
   \label{fig2}
\end{figure}

\subsection{Mobile and immobile densities}

Our aim is to show that the mapping $R^{\ell,\tau}$, defined by Eq.~\eqref{rlt}, converges to a fractional integral, whereas in Appendix~\ref{C} we prove that the quantity $E^{\ell,\tau}$ can be neglected, provided that $P_m^{\ell,\tau}$ and $f^{\ell,\tau}$ converge sufficiently smoothly when $\ell,\tau \to 0$. Indeed, the mapping $R^{\ell,\tau}$ combines convolutions in time and space, with kernels $\tau ^{-1}\Psi_\gamma$ and $\varphi_\ell$. The space convolution with kernel $\varphi_\ell$ converges to the $\text{Id}$ operator~\cite{Sa}. The time convolution with kernel $\tau ^{-1}\Psi_\gamma$ splits into the sum of a singular term, and a mapping that vanishes when $\tau \to 0$. Indeed, hypothesis $H_1$ imposes
\begin{equation}
\tau^{-1}\Psi_\gamma=\lambda t^{-\gamma}/\Gamma(1-\gamma)+\tau^{-1} K(t/\tau^{1/\gamma}),
\end{equation}
$K$ being an integrable kernel. Moreover, in view of Appendix~\ref{B}, the time convolution of kernel $\lambda t^{-\gamma}/\Gamma(1-\gamma)$ is precisely the fractional integral $\lambda I_{0,+}^{1-\gamma}$. For the second term, we have $\Vert \tau^{-1} K(t/\tau^{1/\gamma}) \Vert_{L^{1}(R+)}= \tau^{1/\gamma-1}\Vert K\Vert_{L^{1}(R+)}$, hence Young's inequality (see Appendix~\ref{C}) implies that the convolution of kernel $ \tau^{-1} K(\cdot/\tau^{1/\gamma})$ is a mapping of $L^{1}([0,T],X)$ that vanishes when $\tau\to 0$ for $0<\gamma<1$. Therefore, recollecting the previous results, in the hydrodynamic limit we have
\begin{equation}
P_i(x,t)=\lambda h(x) I_{0,+}^{1-\gamma} P_m(x,t),
\label{Pi}
\end{equation}
and
\begin{equation}
P_m(x,t) = {\mathcal H}_{\lambda,\gamma,h} P(x,t).
\label{Pm}
\end{equation}
Expressions~\eqref{Pi} and~\eqref{Pm} provide the governing equations for the mobile and immobile particles densities, respectively, at the macroscopic scale. When the pdf $\psi$ satisfies $\psi \sim t^{-\gamma-1}$ with $\gamma>1$, so that it has a finite average, the survival probability $\Psi$ is integrable and we have $\lambda=\int_0^{+\infty}t\psi(t)dt = \int_0^{+\infty} \Psi(t)dt$. Then, the time convolution of the kernel $\tau^{-1}\Psi(t/\tau)$ approximates the identity operator $\text{Id}$ when $\ell \to 0$. Hence, the scaling $t_w=\tau W$ leads to $P_i(x,t)=\lambda h(x) P_m(x,t)$. Moreover, Eq.~\eqref{Pm} still holds with ${\mathcal H}_{\lambda,\gamma>1,h}=1/[1+\lambda h(x)]$. Thus, when the sticking times have a finite average, we recover the standard MIM model~\cite{VieVG}, with a retardation factor $1+\lambda h$ (provided that $h$ is uniform).

\subsection{Particles fluxes}

In Appendix~\ref{D}, we show that the probability current ${\mathcal F}^{\ell,\tau}(x,t)$ can be written as $v(t)P_m^{\ell,\tau}(x,t)+{\mathcal F}^{\ell,\tau}_D(x,t) $, up to an additive contribution that vanishes when $\ell, \tau \to 0$, and ${\mathcal F}_D^{\ell,\tau}(x,t) \to -\partial_x DP_m $. Then, using Eq.~\eqref{Pm} and the definition $P=P_m+P_i$ yields the explicit expression for the total tracers flux
\begin{equation}
{\mathcal F}(x,t)= v{\mathcal H}_{\lambda,\gamma,h} P -\partial_x D{\mathcal H}_{\lambda,\gamma,h} P,
\label{fickfrac}
\end{equation}
which generalizes Eq.~\eqref{fick} to spatially distributed trapping events and variable velocity fields: this expression actually represents the Fick's law, as applied to $P_m$. Combining this equation with mass conservation finally gives Eq.~\eqref{eqmim}.

The consistency of Eqs.~\eqref{Pi} and~\eqref{fickfrac} will be verified by showing that solutions to Eq.~\eqref{eqmim} indeed describe the density of a plume of walkers performing the random walks in Sec.~\ref{mim_model}, and that the associated particles fluxes satisfy Eq.~\eqref{fickfrac}.

\begin{figure}[t]
\centerline{\epsfxsize=9.0cm\epsfbox{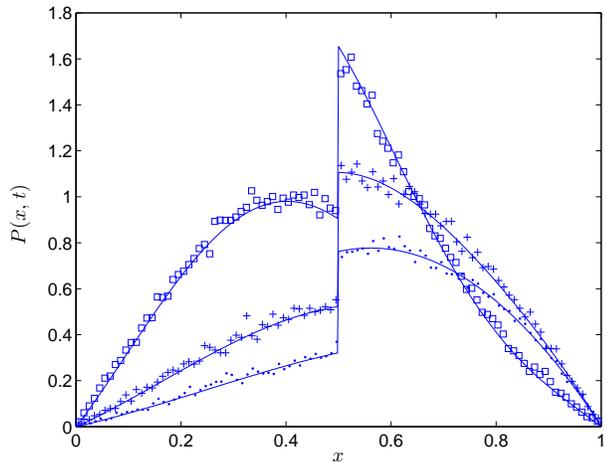}}
\caption{Total tracers concentration $P(x,t)$ at fixed times, on a domain of length $L=1$. In $[0,L/2]$ we have $h(x)=0$, and $h(x)=1$ in $[L/2, L]$. A point-source is located in $x_0=L/4$ at time $t=0$. Both ends of the domain have absorbing boundary conditions. The simulation parameters are: $\gamma=0.8$, $\lambda=1$, $v=0.5$, and $D=0.2$. Curves are plotted at times $t=0.25$ (squares), $t=0.5$ (crosses), and $t=0.75$ (dots).}
   \label{fig3}
\end{figure}

\section{Numerical simulations and comparisons}
\label{montecarlo}

In a previous work, some of the authors discussed the use of numerical schemes discretizing Eq.~\eqref{eqmim} for the case of constant advection field $v$, and unit probability of undergoing a trapping event at the end of each displacement, i.e., $h=1$ \cite{Mary}. In this particular case, Eq.~\eqref{eqmim} is equivalent to the widely adopted Eq.~\eqref{derqmim}, whose Caputo derivatives can be discretized according to various existing numerical schemes \cite{Diethelm, GorAbd}. An alternative integration method was proposed in~\cite{Mary}, so to take advantage of the conservative form of Eq.~\eqref{eqmim}. This scheme can be easily extended to the more general situation addressed here, i.e., the fractal MIM equation~\eqref{eqmim} with time-varying velocity and spatially-dependent sorption probability. Therefore, we proceed now to display numerical solutions of Eq.~\eqref{eqmim}, and to compare them to Monte Carlo simulations of the microscopic-scale random walks described in Sec.~\ref{mim_model}. Indeed, in the hydrodynamic limit, the mobile fraction of an ensemble of random walkers undergoing the stochastic process described in Sec.~\ref{mim_model} approximates the quantity $P_m$, whereas the immobile fraction approximates $P_i$. After briefly revising the essential features of numerical integrations and random walk simulations, we will present comparisons, so to illustrate the theoretical results of Sec.~\ref{equations}, i.e., the Eqs.~\eqref{Pi} and~\eqref{Pm} and the subsequent Eqs.~\eqref{eqmim} and~\eqref{fickfrac}. In particular, we will focus on cases where (although $D$ is uniform) the fractal MIM formulation~\eqref{eqmim} is not equivalent to
\begin{equation}
\left[ \partial_t+\lambda \partial_t^\gamma \right] P(x,t)=\partial_x (\partial_x D-v )P+{\mathcal H}_{\lambda,\gamma,h} r
\end{equation}
which is the version of the more popular fractional differential equation~\eqref{derqmim}, suitable to deal with general source rates $r(x,t)$. Comparisons between pde and random walks were presented in~\cite{Zhang2} for infinite domains; here, we focus exclusively on bounded domains.

\begin{figure}[t]
\centerline{\epsfxsize=9.0cm\epsfbox{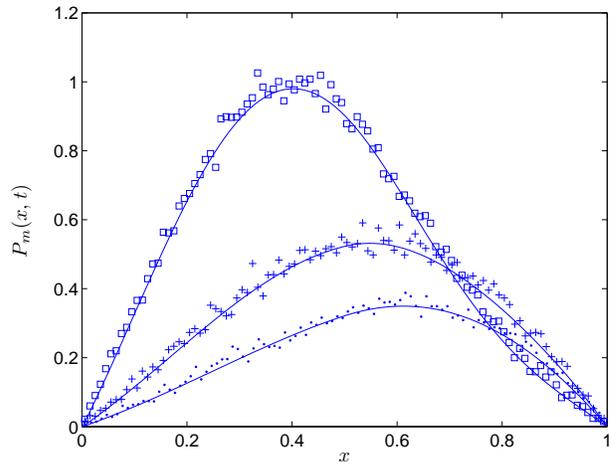}}
\caption{Mobile tracers concentration $P_m(x,t)$ on a domain of length $L=1$, at fixed times, with $h(x)$, initial data and boundary conditions are as in Fig.~\ref{fig3}. The simulation parameters are: $\gamma=0.8$, $\lambda=1$, $v=0.5$, and $D=0.2$. Curves are plotted at times $t=0.25$ (squares), $t=0.5$ (crosses), and $t=0.75$ (dots).}
   \label{fig4}
\end{figure}

\subsection{Numerical methods}

Numerical integration of Eq.~\eqref{eqmim} can be based on an implicit method with centered finite differences schemes for space derivatives, described in~\cite{Mary}, when $P(x,t)$ is smooth, which is the case if $h(x)$ does not show discontinuities. The non-local in time mapping ${\mathcal H}_{\lambda,\gamma,h}$ is approximated by inverting a discrete version of the integral operator $[\text{Id}+\lambda h(x)I_{0,+}^{1-\gamma}]$~\cite{Mary}. Fluxes of tracers are finally given by applying Eq.~\eqref{fickfrac}.

The Monte Carlo particle-tracking approach to the fractal MIM model described above consists in computing the trajectories of a (large) number $N$ of independent particles performing successive displacements, whose rules are defined in Sec.~\ref{mim_model}. More precisely, let us denote by $x_n$ the location after the $n^{th}$ displacement of a particle that originated in $x_0$ at time $t_0$. This walker leaves $x_n$ at time $t_n$, and we have
\begin{equation}
x_{n+1}=x_n+\int_{t_n}^{t_{n}+\tau}v(t')dt'+\sqrt{2D\tau}\xi, \nonumber
\end{equation}
where $\xi$ is a random Gaussian number with zero mean and unit variance, and either
\begin{equation}
t_{n+1}=t_n+\tau+\tau^{1/\gamma}W, \nonumber
\end{equation}
with probability $h(x_{n+1})$, or
\begin{equation}
t_{n+1}=t_n+\tau,
\end{equation}
with probability $1-h(x _{n+1})$.

For the case of constant $v$ and uniform $h=1$, the results in~\cite{Mary} show that random walk simulations are in excellent agreement with numerical integrations of Eq.~\eqref{derqmim}. The same holds for fluxes as in Eq.~\eqref{fickfrac}, obtained from both methods. In the following, we focus our attention on time-dependent velocities, and non-uniform probabilities $h(x)$, in one-dimensional domains $[x_l,x_r]$ with various boundary conditions.

\begin{figure}[t]
\centerline{\epsfxsize=9.0cm\epsfbox{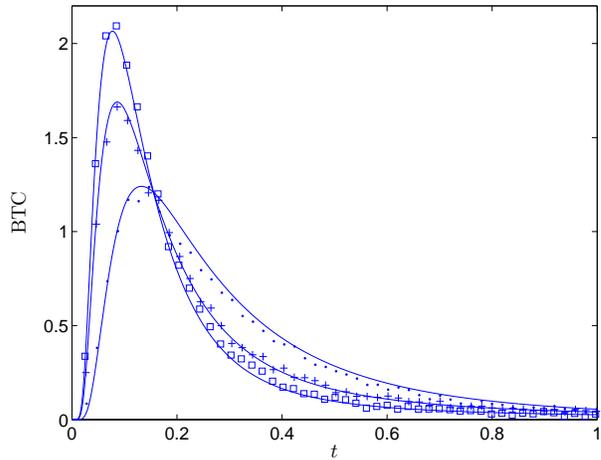}}
\caption{Breakthrough curve ${\mathcal F}(x,t)$ at the right outlet of a domain with length $L=1$. A point-source is located in $x_0=L/4$ at time $t=0$. Both ends of the domain have absorbing boundary conditions. The simulation parameters are: $\lambda=5$, $v=2$, and $D=1$. Symbols represent Monte Carlo simulations ($\gamma=0.8$ squares, $\gamma=0.5$ crosses, $\gamma=0.3$ dots), solid lines the corresponding numerical integration.}
   \label{fig5}
\end{figure}

\subsection{Time-dependent velocity}

We perform comparisons for a periodical velocity $v=\sin{\nu t}$ and $h=1$. In Fig.~\ref{fig1} we display the breakthrough curve (i.e., the outgoing flux from the domain) as a function of time. In the context of solutes transport in porous media, this physical quantity is the most easily accessible by experiments, either at the outlet of laboratory-scale column setups, or at boreholes (wells) for large field-scale measurements. We consider a $1d$ bounded domain of length $L=1$, with a reflective boundary condition at the left, $x_l=0$, and an absorbing boundary condition at the right, $x_r=L$. Velocity is positive pointing towards the right, so that we measure the outgoing flux at the right outlet of the domain. A point source is set at the center of the domain, $x_0=L/2$. An excellent agreement is found between Monte Carlo simulation results and numerically integrated equations.

Then, to further substantiate this analysis, in Fig.~\ref{fig2} we display the spatial concentration profiles $P(x,t)$ for the total tracers concentration (at fixed times). In physical terms, these curves allow quantifying the average displacement and the spread of an initially close plume of injected solutes. Again, a very good agreement is found between Monte Carlo simulations and numerically integrated equations.

\subsection{Non-uniform trapping probability}

In the context of underground contaminant migration, a space-dependent sorption probability $h(x)$ may be expedient to represent the transition between zones of low and high permeability, as well as an alternation of saturated and stagnant regions. Actually, a non-uniform distribution of the trapping sites is expected to be the most common situation in geological formations and complex soils. Conceptually, the simplest case is given by an abrupt variation between $h=0$ and $h=1$ in two adjacent portions of a given domain: this would correspond to the traversed medium being homogeneous and saturated in the former region, and unsaturated and/or heterogeneous in the latter, where retention and sticking effects dominate. Then, the plume migration is Fickian where $h=0$ (memory effects are absent, and transport is ruled by the standard ADE dynamics, as seen from Eq.~\eqref{eqmim}) and anomalous where $h=1$ (transport is ruled by the memory kernel contained in the fractional integral). In the following, we illustrate this case by considering a $1d$ domain of length $L=1$, whose left portion $[0,x_d]$ is characterized by $h=0$, and whose right portion $[x_d, L]$ is characterized by $h=1$. We assume $x_d=L/2$ and set absorbing boundary conditions at both ends of the domain. A point-source is located in $x_0=L/4$ at time $t=0$, i.e., the particles are injected in the homogeneous and saturated region. In Fig.~\ref{fig3} we display the spatial concentration profiles for the total tracers concentration $P(x,t)$ (at fixed times). The most striking feature of this transport process is the appearance of discontinuities in the resident concentration profiles at the interface between the two portions of the domain, whereas the profiles displayed in~\cite{Mary} for $h= 1$ are continuous. This behavior has already been reported elsewhere in the context of the CTRW formulation (see, e.g., the discussions in~\cite{cortis_wrr, hornung, zoia_disc1, zoia_disc2}), and may be understood in terms of the two layers having distinct apparent porosities. In fact, particles experience different trapping times in each layer, and the abrupt variation of the sojourn times distribution at the interface ultimately gives rise to sharp mass accumulation at the interface, as the particles flow is somehow hindered when going from the region with short retention times to the region with long retention times~\cite{cortis_wrr}.

While the implementation of random walks for this case is straightforward, some care is necessary in discretizing densities for numerical integration. More precisely, the total concentration $P(x,t)$ is discontinuous at $x_d$, hence directly discretizing Eq.~\eqref{eqmim} is not convenient. Using instead Eq.~\eqref{Pm}, which links $P$ and $P_m$, and then Fick's law applied to $P_m$, is much more expedient, because $P_m$ is smooth. Combining Eqs.~\eqref{Pm} and~\eqref{eqmim} yields
\begin{eqnarray}
&\partial_t P_m(x,t)={\mathcal H}_{\lambda,\gamma,h}[-\partial_x vP_m+\partial^2_x DP_m+\nonumber \\
&r(x,t)-\lambda h(x)\frac{t^{-\gamma}}{\Gamma(1-\gamma)}P_m(x,0+)]
\label{eqmimPm}
\end{eqnarray}
which is easily discretized following the same lines as in~\cite{Mary}.

In Fig.~\ref{fig4} we display the spatial concentration profiles $P_m(x,t)$ for the mobile tracers concentration (at fixed times). In this case, the curves are smooth across the interface. This is because the particles flux satisfies Fick's law, as applied to $P_m$, and the ADE does not allow for concentration discontinuities at the interface. Indeed, the flux contrasts local variations of $P_m$. Suppose that the profiles of $P$ and $P_m$ have slopes of different signs (e.g., positive for $P_m$ and negative for $P$) at $x$: then, the particles flux through $x$ is negative, and flattens out the spatial increase of $P_m$. On the contrary, the spatial decrease of $P$ has no direct effect on the flux, which therefore does not act on this quantity. For both total and mobile concentrations, a very good agreement is found between Monte Carlo simulations and numerically integrated equations. Finally, Fig.~\ref{fig5} shows the breakthrough curves at the column right outlet, for different values of the exponent $\gamma$ in the absorbing region: again, good agreement is found between Monte Carlo simulations and numerically integrated equations. The asymptotic behavior described by~\cite{Schumer} in infinite domains with $h(x)=  1$ is recovered: even with $h(x)\neq 0$ in some intervals only, $P(x,t)\sim t^{-\gamma}$, $P_m(x,t)\sim t^{-\gamma-1}$ and ${\mathcal F}(x,t)\sim t^{-\gamma-1}$ when $t\to +\infty$. While density profiles obtained for a given value of $\gamma$ with arbitrary $h(x)$ show qualitative differences, fluxes look similar. Hence, BTCs alone are not enough to discriminate between cases.

\section{Conclusions}
\label{conclusions}

In this work, we have discussed a model of contaminant particles flow with trapping events in porous media. Building on the framework of the fractal MIM model, which describes advection-dispersion processes with sticking events in homogeneous flows, we have considered time-varying velocities and space-dependent sticking probabilities. We have first derived the small scale particles dynamics, on the base of a functional relationship between the densities of trapped and non-trapped walkers. This relationship stems directly from the asymptotic behavior of the trapping times distribution, and gives rise to a modified Fick's law with memory for particles fluxes. Then, recalling  mass conservation principle, we have obtained the corresponding governing equations for the evolution of the mobile and immobile phases densities.

These equations have been derived by considering the hydrodynamic (scaling) limit of the underlying microscopic stochastic processes, i.e., by letting the space and time scale of the particles displacements be vanishing small, while preserving the macroscopic dispersion and advection coefficients. The transport equations, which contain non-local in time kernels in the form of fractional integrals, have been discretized and solved numerically by resorting to ad hoc algorithms. Finally, in order to corroborate our results, the contaminant concentration profiles and the breakthrough curves thus obtained by numerical integration have been compared with Monte Carlo particle-tracking simulations.

The relevance and broad applicability of the transport equation~\eqref{eqmim} for the case of non-constant parameters have been emphasized in both theoretical derivations and numerical examples. In particular, we have addressed the case of time-varying velocity fields $v(t)$ and space-dependent trapping probabilities $h(x)$. In fact, the method developed here is more general, and may apply also when the equation coefficients (e.g., $h$) depend on the densities of trapped and mobile walkers, which would result in a nonlinear version of Eq.~\eqref{eqmim}, similarly as in~\cite{LutsBoon, zoia_nonlin}.

Further extensions of our work will address the coexistence of several kinds of traps within the same porous medium, each trap being characterized by a distinct sticking time pdf. This approach would then give rise to slightly more complex mappings ${\mathcal H}_{\lambda,\gamma,h}$ with fractional integrals of distributed order~\cite{ChechGS08}. The simplest case would correspond to two kinds of traps occurring with probabilities $h_1$ and $h_2$, respectively, and sticking time pdfs $\psi_{\gamma_1}$ and $\psi_{\gamma_2}$ (satisfying hypothesis $H_1$ with $\gamma_1<\gamma_2$). Such a model could represent, e.g., multiple phases or regions, with distinct retention properties. Then, the residence times of the solutes would be governed by the mapping ${\mathcal H}_{\lambda_1,\lambda_2,\gamma_1,\gamma_2,h_1,h_2}$, defined as being the inverse of $\text{Id}+\lambda_1 h_1(x)I_{0,+}^{1-\gamma_1}+\lambda_2 h_2(x)I_{0,+}^{1-\gamma_2}$. Moreover, while at intermediate times the solute dynamics is rather involved, at late times the total density would asymptotically decrease as $t^{-\gamma_1}$, i.e., transport would be dominated by the `slower' retention process.

\appendix

\section{Some scenarios for dispersive jumps}
\label{A}

During a given mobile displacement between times $t-\tau$ and $t$, we have a single dispersive jump, whose length is a random variable obeying $\varphi_\ell$. At the scale of microscopic particles trajectories, diverse scenarios may be conceived, which we denote by label $(Si)$. Let $x_{t}^{(i,\ell,\tau)}(\omega)$ be the associated walkers paths.

The source of randomness in each sample $\omega$ arises from the series of successive dispersive jump lengths $J_n$ ($n\geq 1$) and from that of trapping times $T_n$; for convenience, we set $T_n=0$ if there is no trapping event after the $n$-th mobile period. Concerning dispersion, we might consider instantaneous jumps, occurring at the end of each mobile period $(S1)$, at the beginning $(S2)$, or at a random time uniformly distributed in the interval $[t-\tau,t]$ $(S3)$. Alternatively, the dispersive jump might be thought of as being distributed along the total displacement, taking therefore a time $[t-\tau,t]$ to be completed $(S4)$.

Each $\omega$ corresponds to two sequences of numbers, that are the values drawn for the $J_n$ and $T_n$. Of course, these points are identical for all the trajectories $x_{t}^{(i,\ell,\tau)}(\omega)$ started from $x_0$ at time $t=0$. All trajectories pass through points $(t_n+\tau,x_{n+1})$ and $(t_{n+1},x_{n+1})$, with $t_{n+1}=t_n+\tau+T_n$ and $x_{n+1}=x_n+J_n+\int_{t_n}^{t_n+\tau}v(\theta)d\theta$: trapping periods correspond to segments beginning at point $(t_n+\tau,x_{n+1})$, and ending at $(t_n+\tau+T_n,x_{n+1})$, that are common to all scenarios. Hence, the immobile walkers density $P_i^{\ell,\tau}(x,t)$ does not depend on the scenario. Moreover, $\vert x_{t}^{(i,\ell,\tau)}(\omega)- x_{t}^{(j,\ell,\tau)}(\omega)\vert =0$ when $t$ belongs to a trapping period, and $\vert x_{t}^{(i,\ell,\tau)}(\omega) -x_{t}^{(j,\ell,\tau)} (\omega) \vert < \vert J_n\vert$ when $t$ belongs to the $n$-th mobile period. Jumps $J_n$ obey $\varphi_\ell$, and (weakly) converge to zero when $\ell \to 0$. Hence, theorem $3.1$ of~\cite{Billing} implies that, if one among the possible processes $x_{t}^{(i,\ell,\tau)}$ converges to the scaling limit $x_t$ when $\ell \to 0$, then the same holds for all the other paths $x_{t}^{(i,\ell,\tau)}$.

Hence, it follows that the walkers density $P(x,t)$ does not depend on the specific scenario in the hydrodynamic limit, nor does the immobile walkers concentration $P_i(x,t)$. Then, the mobile $P_m$ and immobile densities $P_i$ do not depend on the scenario.

\section{Fractional integrals and derivatives}
\label{B}

The fractional integral $I_{0,+}^{\alpha} f$ of order $\alpha >0$ is
\begin{equation}
I_{0,+}^{\alpha}f(t)=\frac{1}{\Gamma(\alpha)}\int_0^t(t-t')^{\alpha-1}f(t')dt', \nonumber
\end{equation}
which generalizes the usual multiple integrals to non-integer order~\cite{Rub, Sa}. Observe that ${ I}_{0,+}^{\alpha}$ is bounded in $L^p[0,T]$ for $1\leq p\leq\infty$~\cite{Sa}.

The Caputo fractional derivative $\partial_t^\alpha f$ of order $n<\alpha<n+1$ appearing in Eq.~\eqref{derqmim} is defined by
\begin{equation}
\partial_t^\alpha f(t)=I_{0,+}^{n+1-\alpha}\partial^{n+1}_t f(t), \nonumber
\end{equation}
$n$ being an integer~\cite{Sa, MaiUd, Kilb}.

\section{Estimates and limits}
\label{C}

In this Appendix, we will prove that $E^{\ell,\tau} \to 0$ when $\ell,\tau \to 0$, provided that $P_m$ and $f^{\ell,\tau}$ converge. We will make use of some technical results, which will also be used later in Appendix~\ref{D} for fluxes.

\subsection{Hypotheses}

We will need some regularity assumptions for the source rate $r$, the velocity $v$ and the densities $P_m^{\ell,\tau}$ and $f^{\ell,\tau}$.\\

{\em Hypothesis $H_2$}) $r(x,t)$ is the time derivative of some function $\rho(x,t) \in L^{1}(R^+,X)$, i.e., $r(x,t)=\partial_t \rho(x,t)$, and $v$ is uniformly continuous.\\

Observe that initial data of the kind $P(x,0+)=P_0(x)$ lie within this assumption, with $r(x,t)=\delta(t)P_0(x)$.\\

{\em Hypothesis $H_3$.} $i)$ when $\ell,\tau \to 0$ with $D=\ell^2/2\tau$, the density $P_m^{\ell,\tau}$ converges to $P_m$ in $L^1([0,T],X)$, and $\partial_xP_m$ belongs to this space. $ii)$ The distribution $f^{\ell,\tau}$ is the time derivative of some  $F^{\ell,\tau}$ that belongs to $L^1([0,T],X)$: $f^{\ell,\tau}=\partial_t F^{\ell,\tau}$, and $F^{\ell,\tau}\to F$ in $L^1([0,T],X)$ when $\ell,\tau\to 0$, with $f=\partial_t F$. Moreover, $iii)$ $\partial_x (F^{\ell,\tau}+\rho)$ tends to $\partial_x (F+\rho)$ in $L^1([0,T],X)$. Observe that point $iii)$ is not needed if $v$ is constant.\\

\subsection{Statements}

We will make use of Young's inequality, which we reproduce here for convenience~\cite{Arendt}:\\

{\em Young's inequality.} Let $1\leq q',q,q''\leq\infty$, with $1/q'+1/q''=1+1/q$. Then, for $F \in L^{q'}[0,T]$ and $G \in L^{q''}(R)$, we have
\begin{equation}
\Vert F \star G\Vert_{L^q(R)}\leq \Vert F \Vert_{L^{q'}[0,T]}  \Vert  G\Vert_{L^q{q''}(R)}, \nonumber
\end{equation}
and for $F \in L^{q'}(R)$ and $G \in L^{q''}([0,T],{\cal Q})$ we have
\begin{equation}
\Vert F \star G\Vert_{L^q([0,T],{\cal Q})}\leq \Vert F \Vert_{L^{q'}R} \Vert  G\Vert_{L^{q''}([0,T],{\cal Q})}, \nonumber
\end{equation}
where ${\cal Q}$ is a Banach space.\\

We will prove the following proposition:\\

{\em Proposition 1.} Suppose that hypotheses $H_1$, $H_2$ and $H_3$ are satisfied. Then, $i)$ $\Psi_\gamma * \varphi_\ell \star  \varepsilon^{\ell,\tau} \to 0$ in $L^1([0,T],X)$. Moreover, $ii)$ $[H\Phi(\frac{\cdot}{\ell})]\star \varepsilon^{\ell,\tau} \to 0$ in the set ${\mathcal S}'([0,T],X)$ of tempered distributions.\\

Note that point $i)$ implies $E^{\ell,\tau} \to 0$ in the hydrodynamic limit. The proof will use the following lemmas.\\

{\em Lemma 1.} $i)$ Let $w$ be a continuous function. Then, for $g$ in $L^1([0,T],X)$, ${\mathcal T}_{t'}{\mathcal Y}_{w(t')}g$ is a continuous function of $t'$, with values in $L^1([0,T],X)$. Moreover, $ii)$ the mapping ${\mathcal T}_{t'}{\mathcal Y}_{w(t')}$ is a contraction in $L^1([0,T],X)$.\\

{\em Lemma 2.} If $g \in L^1([0,T],X)$, then $\int_0^1 [{\mathcal T}_{\tau}{\mathcal  Y}_{L(t,\tau}-{\mathcal T}_{\theta\tau} {\mathcal Y}_{L(t,\theta\tau)}]g(x,t)d\theta \to 0$ in $L^1([0,T],X)$, when $ \tau \to 0$.\\

{\em Consequence.} If hypotheses $H_2$ and $H_3$ are satisfied, then $\int_0^1 [{\mathcal T}_{\tau}{\mathcal  Y}_{L(t,\tau)}-{\mathcal T}_{\theta\tau}{\mathcal  Y}_{L(t,\theta\tau)}] [F^{\ell,\tau} + \rho](x,t)d\theta \to 0$ in $L^1([0,T],X)$ when $\tau \to 0$.

\subsection{Proofs}

{\em Proof of Proposition 1.} Due to $H_2$ and $H_3$, we have
\begin{eqnarray}
&\varepsilon^{\ell,\tau}(x,t)=\nonumber \\
&\partial_t\int_0^1 [{\mathcal T}_{\tau}{\mathcal  Y}_{L(t,\tau)}-{\mathcal T}_{\tau\theta} {\mathcal Y}_{L(t,\tau\theta)}] (F^{\ell,\tau}+\rho)d\theta+I_1,
\label{B1}
\end{eqnarray}
where
\begin{eqnarray}
&I_1=\int_0^1[(v(t)-v(t-\tau)){\mathcal  T}_{\tau}{\mathcal  Y}_{L(t,\tau)}-\nonumber \\
&(v(t)-v(t- \theta\tau)){\mathcal T}_{\tau\theta}{\mathcal  Y}_{L(t,\tau\theta)}] \partial_x (F^{\ell,\tau}+\rho)d\theta.\nonumber
\end{eqnarray}
Therefore, we obtain
\begin{eqnarray}
&\Psi_\gamma * \varepsilon^{\ell,\tau}=\psi_\gamma *\int_0^1 [{\mathcal T}_{\tau}{\mathcal  Y}_{L(t,\tau)}-\nonumber \\
&{\mathcal T}_{\tau\theta}{\mathcal  Y}_{L(t,\tau\theta)}][ F^{\ell,\tau}+\rho]d\theta+\Psi_\gamma *I_1.
\label{B2}
\end{eqnarray}
The second integral $I_1$ on the r.~h.~s.~of Eq.~\eqref{B1} vanishes if $v$ is constant. If $v$ is not constant, it tends to zero when $\tau \to 0$ in $L^1([0,T],X)$, in view of $H_2$ and $H_3$ $iii)$. Then, since $\Psi_\gamma(t)$ is bounded (by $1$), $\Psi_\gamma*I_1\to 0$ in $L^\infty ([0,T],X)$, hence in $L^1([0,T],X)$ since $T$ is finite.  Moreover, since $\psi_\gamma$ is normalized, the Consequence and Young's inequality imply that the first term on the of r.~h.~s.~of Eq.~\eqref{B2} vanishes, which proves point $i)$.\\

For non-constant $v$, hypothesis $H_3$ $iii)$ implies that
\begin{eqnarray}
&\Phi(\cdot/\ell)\star I_1=\varphi_\ell\star\int_0^1[(v(t)-v(t-\tau)){\mathcal  T}_{\tau}{\mathcal  Y}_{L(t,\tau)}- \nonumber \\
&(v(t)-v(t- \theta\tau)){\mathcal T}_{\tau\theta}{\mathcal  Y}_{L(t,\tau\theta)}] [f^{\ell,\tau}+\rho](x,t)d\theta \to 0 \nonumber
\end{eqnarray}
in $L^1([0,T],X)$, since $I_1\to 0$ while  $\varphi_\ell$ is normalized. Besides, ${\mathcal U}=\Phi(\cdot/\ell)\star\partial_t\int_0^1 ({\mathcal T}_{\tau}{\mathcal  Y}_{L(t,\tau)}-{\mathcal T}_{\tau\theta}{\mathcal  Y}_{L(t,\tau\theta)}) (F^{\ell,\tau}+\rho)(x,t)d\theta$ may not belong to $L^1([0,T],X)$. Nevertheless, ${\mathcal U}$  is the time derivative of $\Phi(\cdot/\ell)\star\int_0^1 ({\mathcal T}_{\tau}{\mathcal  Y}_{L(t,\tau)}-{\mathcal T}_{\tau\theta}{\mathcal  Y}_{L(t,\tau\theta)}) [F^{\ell,\tau}+\rho](x,t)d\theta$, which vanishes in this space. This proves point $ii)$.\\

We have now to prove the Lemmas. In Lemma $1$, point $i)$ follows from the proof of Theorem $9.5$ of~\cite{Rudin}, stating that ${\mathcal Y}_aG$ is a continuous function of $a$, with values in $L^1(R)$, provided that we have $G \in L^1(R)$. Point $ii)$ is obvious.

For Lemma $2$, the function ${\mathcal T}_{\theta\tau}{\mathcal  Y}_{L(t,\theta\tau)}g(x,t)$ belongs to $L^1([0,T],X)$, and depends continuously on $\theta$ by Lemma $1$ $i)$. Hence, it is Bochner-integrable~\cite{Arendt} from $[0,1]$ to $L^1([0,T],X)$. Moreover, by Lemma $1$ $ii)$, ${\mathcal T}_{\theta\tau}{\mathcal  Y}_{L(t,\theta\tau)}g(x,t) \to g(x,t)$ pointwise, whereas $\Vert{\mathcal  T}_{\theta\tau}{\mathcal  Y}_{L(t,\theta\tau)}g(x,t)-g\Vert_Y \leq \Vert g\Vert _Y$ in  $Y=L^1([0,T],X)$ norm, so that dominated convergence proves the Lemma.

Finally, the Consequence is immediate from Lemma $2$, since we have $\int_0^1 ({\mathcal T}_{\tau}{\mathcal  Y}_{L(t,\tau)}-{\mathcal  T}_{\theta\tau}){\mathcal  Y}_{L(t,\theta\tau)}[F+\rho](x,t)d\theta \to 0$ in $L^1([0,T],X)$ by Lemma $2$, and $\int_0^1 ({\mathcal T}_{\tau}{\mathcal  Y}_{L(t,\tau)}-{\mathcal  T}_{\theta\tau}{\mathcal  Y}_{L(t,\theta\tau)})[F^{\ell,\tau}-F](x,t)d\theta \to 0$ due to $H_3$ and Lemma $1$.

\section{Probability current}
\label{D}

\subsection{The Fickian case}
\label{f1}

The particles flux (probability current) ${\mathcal F}^{\ell,\tau}(x,t)$ of the process $x_t^{\ell,\tau}$ represents the average net number of walkers crossing point $x$ at time $t$. Its dispersive and advective components are
\begin{equation}
{\mathcal F}^{\ell,\tau}(x,t)={\mathcal F}_D^{\ell,\tau}(x,t)+v(t)P^{\ell,\tau}(x,t),
\label{ade_flux}
\end{equation}
where ${\mathcal F}_D^{\ell,\tau}(x,t)$ denotes the contribution of dispersive jumps, and $P^{\ell,\tau}(x,t)$ is the density of the process $x_t^{\ell,\tau}$. Since each walker performs one dispersive jump per time step $\tau$, we have
\begin{equation}
{\mathcal F}_D^{\ell,\tau} =\int_0^{+\infty}\frac{P^{\ell,\tau}(x-y,t)-P^{\ell,\tau}(x+y,t)}{\tau}\Phi(\frac{y}{\ell})dy,
\label{disp_flux}
\end{equation}
where the function $\Phi(y/\ell)= \int_{y/\ell}^{+\infty}\varphi_\ell(z)dz$ represents the probability that dispersive jump length is larger than $y$. Then, recalling that $\varphi_\ell$ is symmetric, we can rewrite ${\mathcal F}_D^{\ell,\tau}(x,t)={\mathcal F}_{D,+}^{\ell,\tau}(x,t)-{\mathcal F}_{D,-}^{\ell,\tau}(x,t)$, where
\begin{equation}
{\mathcal F}_{D,\pm}^{\ell,\tau}=2\int_0^{+\infty}\frac{(DP^{\ell,\tau})(x\mp y,t)-(DP^{\ell,\tau})(x,t)}{\ell^2}\Phi(\frac{y}{\ell})dy,
\label{disp_fluxpm}
\end{equation}
with $D=\ell^2/2\tau$. Hence, the contribution of dispersive jumps to the probability current is expressed through convolutions, whose kernel has a form $\ell^{-\alpha-1}{\mathcal K}(y/\ell)$. Under some assumptions on ${\mathcal K}$, such mappings have a limit when $\ell \to 0$, that is a derivative of order $\alpha$~\cite{Sa1, Sa2, MeerSchef1, Neel8}. For the case considered here, $\alpha=1$, the lemma below shows that ${\mathcal F}_D^{\ell,\tau}$ converges in the hydrodynamic limit to $-\partial_x DP$. The lemma applies to Eq.~\eqref{disp_fluxpm} if $DP$ has a uniformly bounded derivative with respect to $x$, provided that also $P^{\ell,\tau}-P\to 0$ more rapidly than $\ell$. In domains limited by reflecting boundaries, the dispersive flux needs corrections with respect to Eq.~\eqref{disp_flux} on the small scale, due to particles bouncing back at the walls~\cite{Neel}. Nevertheless, Eq.~\eqref{fick} still holds for symmetric $\varphi_\ell$ with a finite second moment. Also, care must be taken when dealing with absorbing boundary conditions if $\varphi_\ell$ has a diverging second moment~\cite{zoia_levy}.\\

{\em Lemma 3.} Let $\Phi$ be a differentiable function, integrable over $R^+$, positive and decreasing. Then, for any integrable function $g$ whose derivative is uniformly bounded,
\begin{equation}
\int_0^{+\infty}\frac{g(x+\ell y)-g(x)}{\ell}\Phi(y)dy \to -\frac{1}{2}\frac{dg}{dx}\int_{0}^{+\infty}\Phi'(y)y^2dy \nonumber
\end{equation}
pointwise when $\ell\to 0$.\\

This proposition appears in~\cite{Neel}, within a slightly different context. Moreover, since $\Phi(y)=\int_y^{+\infty}\varphi_1(z)dz$, we have $\Phi'(y)=-\varphi_1(y)$.\\

{\em Proof.} Let us denote ${\mathcal A}(\ell)$ a function, such that ${\mathcal A}(\ell) \to +\infty$ when $\ell \to 0$, with $\ell{\mathcal A}(\ell) \to 0$. For instance, we can assume ${\mathcal A}(\ell)=\ell^{-a}$, with $0<a<1$. Then, we have
\begin{equation}
\int_0^{+\infty}\frac{g(x+\ell y)-g(x)}{\ell}\Phi(y)dy, \nonumber
\end{equation}
which can be written as ${\mathcal I}(\ell)+ {\mathcal J}(\ell)$, with
\begin{equation}
{\mathcal I}(\ell)=\int_0^{{\mathcal A}(\ell)}\frac{g(x+\ell y)-g(x)}{\ell y}y\Phi(y)dy \nonumber
\end{equation}
and
\begin{equation}
{\mathcal J}(\ell)=\int_{{\mathcal A}(\ell)}^{+\infty}\frac{g(x+\ell y)-g(x)}{\ell y }y\Phi(y)dy.\nonumber
\end{equation}
Then, ${\mathcal I}(\ell) \to \frac{dg}{dx}\int_{0}^{+\infty}y\Phi(y)dy$, which integrating by parts yields $\int_{0}^{+\infty}y\Phi(y)dy = -\frac{1}{2} \int_{0}^{+\infty}\Phi'(y)y^2dy$.

Finally, for ${\mathcal J}(\ell)$ we have $\vert {\mathcal J}(\ell)\vert\leq M\vert \int_{{\mathcal A}(\ell)}^{+\infty}\Phi(y)dy \vert$, with $M$ denoting the sup norm of the derivative of $g$. This term vanishes in the hydrodynamic limit, in view of the above choice of ${\mathcal A}(\ell)$.

\subsection{The case of dispersion with immobile periods}
\label{f2}

An explicit expression for the particles flux can be derived also for random walks with immobile periods. We will do it within scenario $(S1)$. We further assume that $ P_m^{\ell,\tau}$ and $ f^{\ell,\tau}$ converge according to hypothesis $H_3$ of Appendix~\ref{C}, on the basis of the probability for a tagged particle to cross $x$ to the left/right during a small time interval.

Particles that cross point $x$ towards the right during time interval $[t,t+dt]$ must be mobile, and have spent a time $t'\in[0,\tau]$ in the mobile period. Moreover, they may $i)$ or may not $ii)$ have completed the single dispersive jump. In the former case $i)$, they spent exactly a time $\tau$ in that period, that began at point $x-y-L(t,\tau)$ between instants $t-\tau$ and $t+dt-\tau$, if the jump length is larger than $y$. Collecting all positive contributions $y>0$ gives for case $i)$ the probability
\begin{equation}
\int_0^{+\infty} {\mathcal  T}_{\tau}{\mathcal Y}_{L(t,\tau)} [f^{\ell,\tau}+r](x-y,t)\Phi(\frac{y}{\ell})dydt. \nonumber
\end{equation}
Case $ii)$ can not occur if $v(t)<0$. For positive $v(t)$, it corresponds to particles that entered the mobile period between points $x-v(t)dt-L(t,t')$ and $x-L(t,t')$, for all values of $t'\in[0,\tau]$, which yields the probability
\begin{equation}
\int_0 ^\tau {\mathcal T}_{t'}{\mathcal Y}_{L(t,t')} [f^{\ell,\tau}+r](x,t)dt'v(t)dt. \nonumber
\end{equation}
Upon dividing by $dt$, we recognize $v(t)P_m^{\ell,\tau}(x,t)$, according to Eq.~\eqref{pmlt}. If $v(t)>0$, crossings towards the left correspond to dispersive jumps and we only have
\begin{equation}
\int_0^{+\infty}{\mathcal  T}_{\tau}{\mathcal Y}_{L(t,\tau)} [f^{\ell,\tau}+r](x+y,t)\Phi(\frac{y}{\ell})dydt \nonumber
\end{equation}
Hence, in view of Eq.~\eqref{epslt} the probability current will be given by the sum of two terms
\begin{equation}
\int_0^{+\infty}\frac{P_m^{\ell,\tau}(x-y,t)-P_m^{\ell,\tau}(x+y,t)}{\tau}\Phi(\frac{y}{\ell})dy+v(t)P^{\ell,\tau}_m(x,t),\nonumber
\end{equation}
which tends to $-\partial_x DP_m+v(t)P_m$ as shown above, and
\begin{equation}
\int_0^{+\infty}\varepsilon^{\ell,\tau}(x - y,t)\Phi(\frac{y}{\ell})dy - \int_0^{+\infty}\varepsilon^{\ell,\tau}(x + y,t)\Phi(\frac{y}{\ell})dy. \nonumber
\end{equation}
This latter expression vanishes, according to the proposition $1$ of Appendix~\ref{C}, which proves Eq.~\eqref{fickfrac}.

\end{document}